\documentclass[aps,prb,twocolumn]{revtex4}
\usepackage{amsmath,amssymb}
\usepackage{graphicx}
\usepackage{times}
\usepackage{hyperref}

\begin{document}
\title{Fragile magnetic order in the honeycomb lattice Iridate Na$_{2}$IrO$_3$ revealed by magnetic impurity doping}
\author{Kavita Mehlawat and Yogesh Singh}
\affiliation{Indian Institute of Science Education and Research (IISER) Mohali, Knowledge city, Sector 81, Mohali 140306, India}

\date{\today}

\begin{abstract}
We report the structure, magnetic, and thermal property measurements on single crystalline and polycrystalline samples of Ru substituted honeycomb lattice iridate Na$_2$Ir$_x$Ru$_{1-x}$O$_3$ (x = 0, 0.05, 0.1, 0.15, 0.2, 0.3, 0.5).  The evolution of magnetism in Na$_2$Ir$_x$Ru$_{1-x}$O$_3$ has been studied using dc and ac magnetic susceptibility, and heat capacity measurements.  The parent compound Na$_2$IrO$_3$ is a spin-orbit driven Mott insulator with magnetic order of reduced moments below $T_N = 15$~K\@.  In the Ru substituted samples the antiferromagnetic long range state is replaced by a spin glass like state even for the smallest substitution suggesting that the magnetic order in Na$_2$IrO$_3$ is extremely fragile.  We argue that these behaviors indicate the importance of nearest-neighbor magnetic exchange in the parent Na$_2$IrO$_3$.  Additionally, all samples show insulating electrical transport.  
\end{abstract}

\maketitle
\section{Introduction}
\label{sec:INTRO}
Transition metal oxides (TMO) display a large variety of novel physical phenomena arising from the mutual interplay of coulomb correlation (U), bandwidth (W), and spin-orbit coupling ($\lambda $).  In recent years, there has been an increased interest in the study of 5d transition metal oxides where these energy scales are all comparable.  Iridium based oxides such as the square-lattice materials (Sr,Ba)$_{2}$IrO$ _{4}$ ~\cite{Kim2009, Kim2008, Okabe2011}, the perovskites SrIrO$_3$ and Sr$_3$Ir$_2$IO$_7$ ~\cite{Moon2008}, the pyrochlore materials A$_{2}$Ir$ _{2} $O$ _{7} $ (A = Y,Eu,Lu,Sm) ~\cite{Gardner2010, Yang2011, Zhao2011}, the honeycomb-lattice materials A$_{2}$IrO$ _{3} $ (A = Na,Li) ~\cite{Jackeli2009, Chaloupka2010, Singh2010}, and the hyperkagome-lattice material Na$_{4}$Ir$ _{3} $O$ _{8}$ ~\cite{Okamoto2007, Singh2013} have been materials of intensive recent investigations.  The $5d$ transition metal oxides are expected to be wide-band weakly correlated metals due to large spatial extent of their $d$-orbitals.  However, the above mentioned iridates belong to a new class of spin-orbit assisted Mott insulators where the insulating state arises from the combined effect of strong spin-orbit coupling and coulomb correlation ~\cite{Kim2008, Pesin2010}. 

In the present study, we focus on the honeycomb lattice iridate Na$_{2}$IrO$ _{3} $ in which spin-orbit entangled effective moments $J_{eff} = 1/2$ sit on a honeycomb lattice leading to interesting magnetic properties \cite{Singh2010} which have been discussed in relation to the Kitaev-Heisenberg model \cite{Jackeli2009, Chaloupka2010} and various extensions.  Na$_{2}$IrO$ _{3} $ is electrically insulating with a measured band gap of 350~meV ~\cite{Comin2012}.  The magnetic susceptibility reveals local moment character with effective $S = 1/2$ moments with predominant antiferromagnetic coupling as indicated by a large and negative Weiss temperature $\theta$ = -120 K.  Long range antiferromagnetic ordering however, occurs at a much reduced temperature of $T_{N} \approx 15$~K suggesting strong magnetic frustration.  The reduced magnetic entropy recovered above the magnetic transition further suggests reduced moment ordering or partially fluctuating moments below $T_N$ \cite{Singh2010}.  The zig-zag magnetic order observed for Na$_{2}$IrO$ _{3}$ cannot be realized within a nearest-neighbor Kiteav-Heisenberg \cite{Chaloupka2013}, unless ferromagnetic (FM) Heisenberg and antiferromagnetic (AF) Kiteav exchange couplings are used \cite{Chaloupka2013, Choi2012}.  Alternatively an extended Heisenberg-Kitaev model including further-neighbor interactions can give the zig-zag order \cite{Singh2012}. The presence of such further neighbor exchanges has also been suggested from the analysis of the observed low energy magnon dispersion in Na$_{2}$IrO$ _{3}$ \cite{Choi2012}.  

It is of interest to ask how the novel properties of Na$_2$IrO$_3$ will change under various perturbations like charge doping, externally applied pressure, etc.  Recently there has been a study on the non-magnetic dilution of $A_2$IrO$_3$ ($A =$ Na, Li) by the partial replacement of Ir by non-magnetic Ti.  This study revealed the importance of near-neighbor exchange in Na$_2$IrO$_3$ while in Li$_2$IrO$_3$ further than nerest-neighbor interactions were found to be consistent with their observations of a magnetic or spin-glass temperature which was suppressed at the percolation limit in the Na case but which persisted beyond this limit for the Li system \cite{Manni2014}.  It has been predicted that a superconducting ground state would emerge with hole doping in the Kitaev-Heisenberg model \cite{You2012, Okamoto2013}.  

In this work we explore how the properties of Na$_2$IrO$_3$ evolve when Ir is partially replaced by a magnetic ion of a different spin (Ru).  Our particular interest is to understand how the structural, magnetic, transport, and thermal properties change due to magnetic impurity doping.  To this end we have synthesized single crystals and polycrystals of Na$_2$Ir$_x$Ru$_{1-x}$O$_3$ (x = 0, 0.05, 0.1, 0.15, 0.2, 0.3, 0.5) and studied their crystal structure, electrical transport, dc magnetic susceptibility, ac magnetic susceptibility, and heat capacity.  

Beyond $x = 0.5$ (specifically $x = 0.65, 0.75$ were attempted) we obtain mixed phase samples with Na$_2$IrO$_3$ and Na$_2$RuO$_3$ phase sepaerated. Attempts to synthesize Na$_2$RuO$_3$ were not successful.  Rather suprisingly we find that all samples ($x \leq 0.5$) remain insulating with Ru contributing localized moments to the magnetism.  Additionally the long-ranged magnetic order of the parent Na$_2$IrO$_3$ is replaced by a frozen spin-glassy state even for the smallest Ru substitution, suggesting an extremely fragile magnetic order and the presence of several competing magnetic states.    

\section{EXPERIMENTAL DETAILS}
\label{sec:EXPT}
The single crystalline ($x = 0, 0.1, 0.15, 0.20, 0.30$) and single phase polycrystalline ($x = 0, 0.05, 0.10, 0.30, 0.50$) samples of Na$_2$Ir$_x$Ru$_{1-x}$O$_3$ were synthesized.  The starting materials were Na$_2$CO$_3$ (99.995$\%$ Alfa Aesar), Ru powder (99.95$\%$ Alfa Aesar) and anhydrous IrO$ _{2} $(99.95$\%$ Alfa Aesar) or Ir metal powder (99.95$\%$ Alfa Aesar).  Single crystals were grown using a self flux growth method using off-stoichiometric amounts of starting materials as described previously elsewhere \cite{Singh2010} and polycrystalline samples were synthesized by using standard solid state reaction methods as described in detail previously elsewhere \cite{Singh2010}.  Plate like crystals were found vertically standing over a polycrystalline platform, randomly stacked and attached with one another at random angles.  The structure and composition of the resulting samples were checked by using powder x-ray diffraction (PXRD) and chemical analysis using energy dispersive x-ray (EDX) analysis with a JEOL scanning electron microscope (SEM).  The PXRD was obtained by a Rigaku diffractometer with Cu K$_{\alpha}$ radiation in 2$ \theta $ range from 10$^\circ$ to 90$^\circ$ with 0.02$^\circ$ step size.  Physical property measurements of electrical transport, dc susceptibility, ac susceptibilty, and heat capacity was done using a Quantum Design physical property measurement system. 

\section{RESULTS}

\subsection{Crystal Structure and Chemical Analysis}

The powder x-ray diffraction patterns of all single phase samples could be indexed with the C2/m space group.  The cell parameters  extracted from single crystal diffraction or from the PXRD data are listed in table~\ref{Table-lattice-parameters}.  A full single crystal refinement was not possible because of the presence of multiple twins rotated around the $c^*$ axis in the crystals measured.  We find that the cell parameters do not change monotonically as increasing amounts of Ru are introduced into the system.  From the cell parameters it can be seen that initially for $x = 0.05, 0.1$, the $a$ and $b$ lattice parameters reduce with increasing Ru content while the $c$-axis parameter does not change appreciably.  Thus, initially the honeycomb lattice shrinks in-plane while the inter-layer separation stays approximately the same as Ru is partially substituted for Ir.  However, for $x = 0.20$, $a$ and $b$ increase again.  For $x = 0.50$ the trend reverses again. From the ionic sizes of Ru$^{4+}$ and Ir$^{4+}$ one expects the volume to shrink when the smaller Ru$^{4+}$ is replacing the larger Ir$^{4+}$.  Therefore the observed non-monotonic trend is difficult to understand.  We note that recently single crystalline Na$_2$RuO$_3$ has been synthesized and its lattice parameters were also found to be larger than those of single crystal Na$_2$IrO$_3$ contrary to expectation from ionic sizes \cite{Wang2014}. 
    
Chemical analysis using energy dispersive spectroscopy on several spots of the same crystal and on several crystals with the same nominal starting composition have been performed.  The average value of $x$ is given in Table~\ref{Table-lattice-parameters} and is compared with the nominal starting composition.  The obtained Ru concentrations are within a few percent of the target Ru content, therefore the nominal $x$ will be used.       

\begin{table}

\caption{Lattice Parameters of Na$_2$Ir$_x$Ru$_{1-x}$O$_3$ from single crystal ($x \approx 0, 0.1, 0.2$) and powder diffraction ($x \approx 0.5$)}

\begin{ruledtabular}

\begin{tabular}{|c|ccccc|}
x & $Space $ $Group$ & $a$ & $b$ & $c$ &   $\beta$ \\ \hline  
0 & C2 /m & 5.43& 9.40 & 5.61 & 109.04  \\
0.1 & C2 /m & 5.39& 9.34 & 5.63 & 108.45  \\
0.2 & C2 /m & 5.42& 9.38 & 5.64 & 108.51  \\ 
0.5 & C2 /m & 5.35& 9.36 & 5.62 & 108.57  \\ 
\end{tabular}

\end{ruledtabular}
\label{Table-lattice-parameters}
\end{table}

\subsection{DC Magnetic susceptibility}
\begin{figure}[t]   

\includegraphics[width= 3 in]{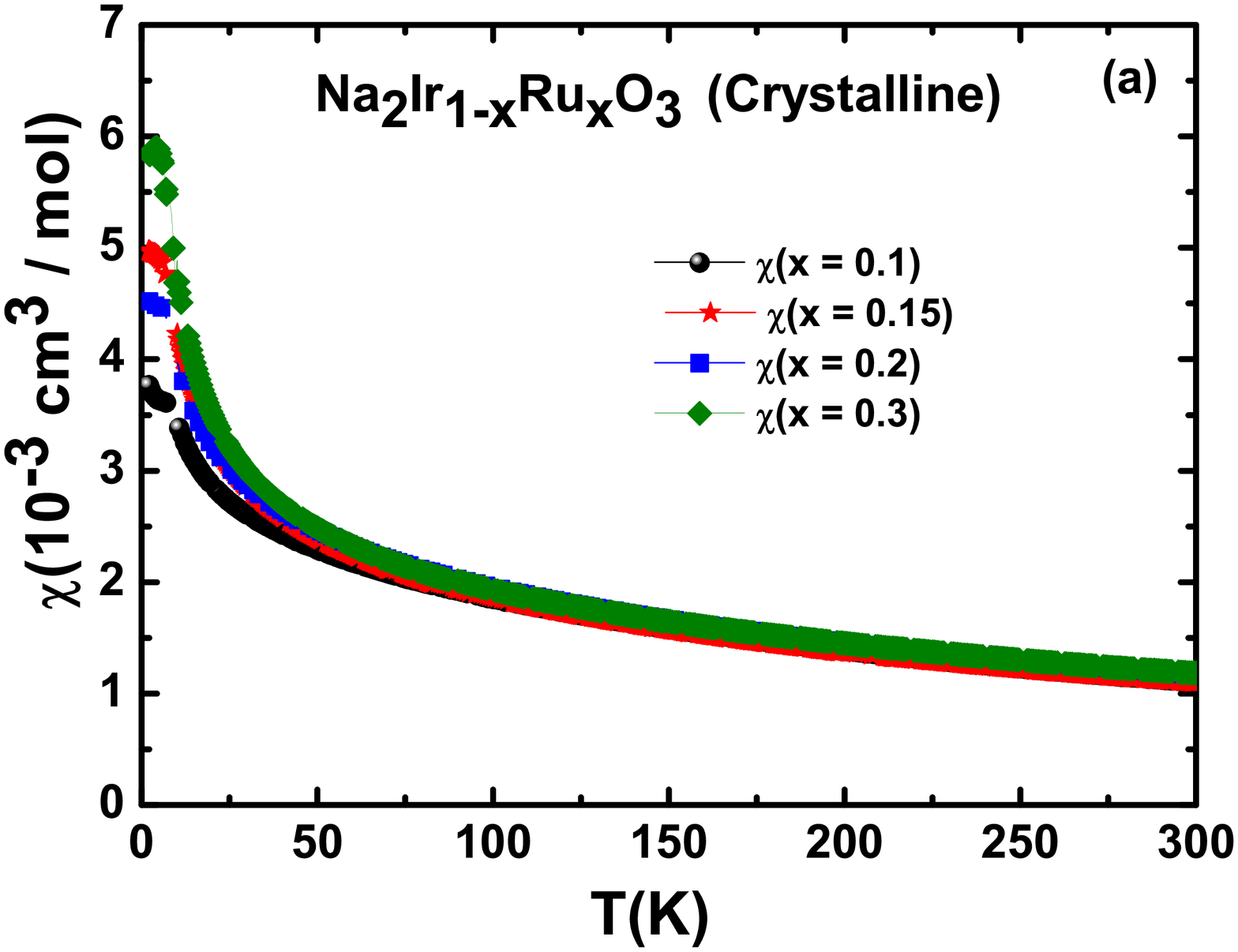}
\includegraphics[width= 3 in]{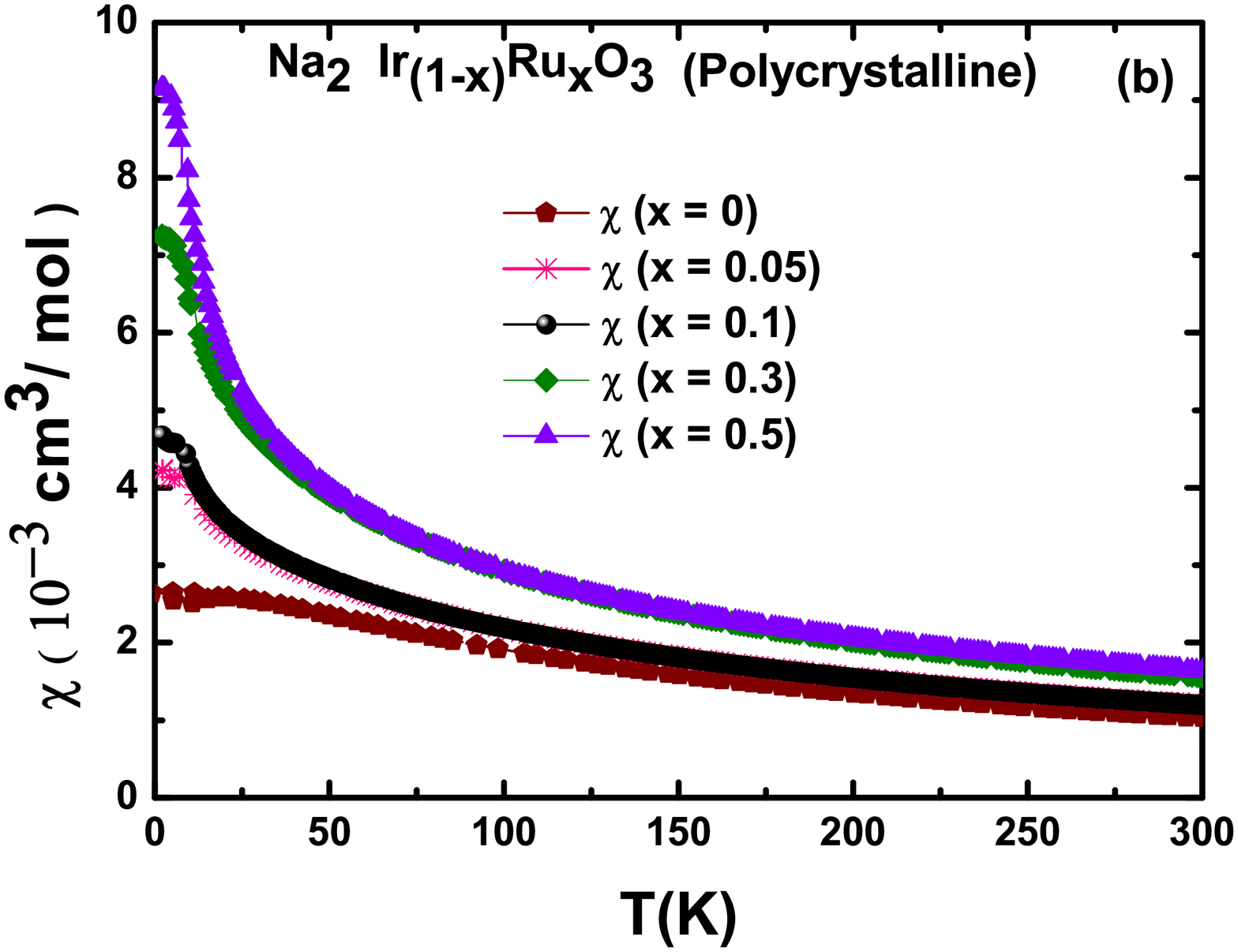}      
\caption{(Color online) Dc magnetic susceptibility $\chi$ versus $T$ of Na$_2$Ir$_x$Ru$_{1-x}$O$_3$ between $T = 2$~K and 305~K\@. (a) Show Dc magnetic susceptibility data of single crystalline Na$_2$Ir$_x$Ru$_{1-x}$O$_3$ ($x \approx  0.1, 0.15, 0.2, 0.3 $) at applied field $ H = 1~Tesla$ (b) Show Dc magnetic susceptibility data of Polycrystalline Na$_2$Ir$_x$Ru$_{1-x}$O$_3$ ($x \approx  0.05, 0.1, 0.2, 0.3 $) at applied field $ H = 1~Tesla $.  
\label{Fig-1}}
\end{figure}

The magnetic susceptibility $\chi = M / H$ versus T data for Na$_2$Ir$_x$Ru$_{1-x}$O$_3$ between $T = 2$~K and $305$~K measured in an applied magnetic field $H = 1$~T are shown in Fig.~\ref{Fig-1}.  Figure~\ref{Fig-1}~(a) shows the $\chi(T)$ data for single crystalline samples ($x = 0.10, 0.15, 0.20, 0.30$).  The field was applied parallel to the $ab$-plane.  Figure~\ref{Fig-1}~(b) shows the $\chi(T)$ data for polycrystalline samples ($x = 0, 0.05, 0.10, 0.30, 0.50$).  All samples show the behavior of local moment magnetism indicating that Ru substitution does not lead to charge carrier doping.  The magnitude of $\chi(T)$ increases with increasing Ru concentration as expected on substituting $S = 1/2$, Ir$^{4+}$ localized moments with $S = 1$, Ru$^{4+}$ localized moments.   

The $\chi(T)$ data between $T = 200$~K and $305$~K for the polycrystalline samples were fit by the Curie-Weiss expression $\chi = \chi_0+{C\over T-\theta}$ where $\chi_0$, $C$, and $\theta$ are the fitting parameters.  The parameters obtained from fits to the data for the samples Na$_2$Ir$_x$Ru$_{1-x}$O$_3$ ($x \approx 0, 0.1, 0.3, 0.5$) are given in Table~\ref{Table-chi}. Assuming a $ g$-factor $g = 2$, the effective moment $\mu_{eff}$ has been estimated from the obtained value of the Curie constant $C$.  These $\mu_{eff}$ values are also listed in Table~\ref{Table-chi} for each sample.  For the parent $x = 0$ compound $\mu_{eff}$ = 1.81(2)~$\mu_{B} $ is close to the previously reported value \cite{Singh2010}.  The value of $\mu_{eff}$ monotonically increases with increasing $x$ indicating that the effective moment increases, as is expected if Ir$^{4+}$ ($S = 1/2$) moments are replaced by Ru$^{4+}$ ($S = 1$) local moments.  The value of the Weiss temperature $ \theta $ stays large and negative indicating persisting strong antiferromagnetic interactions.

\begin{table}

\caption{ Parameters obtained from fits to the magnetic susceptibility data by the Curie-Weiss expression $\chi = \chi_0 + {C \over T - \theta}$ }

\begin{ruledtabular}

\begin{tabular}{|c|cccc|}
x & $\chi_0$~($10^{-5}$~cm$^3$/~mol) & $C$~(cm$^3$~K/~mol) & $\theta$~(K) & $ \mu_{eff} $  $(\mu_{B}) $ \\ \hline  
0 & 3.1(4)& 0.41(7) & -113(1) & 1.81(1)  \\
0.1 & 6.9(1)&0.47(8) & -124(2) & 1.94(2)  \\ 
0.3 & 18.4(6)& 0.56(2) & -105(4) & 2.12(4)  \\ 
0.5 & 17.8(2)& 0.64(4) & -138(1) & 2.26(3) \\ 
\end{tabular}

\end{ruledtabular}
\label{Table-chi}
\end{table} 

\begin{figure}[t]
\includegraphics[width= 3 in]{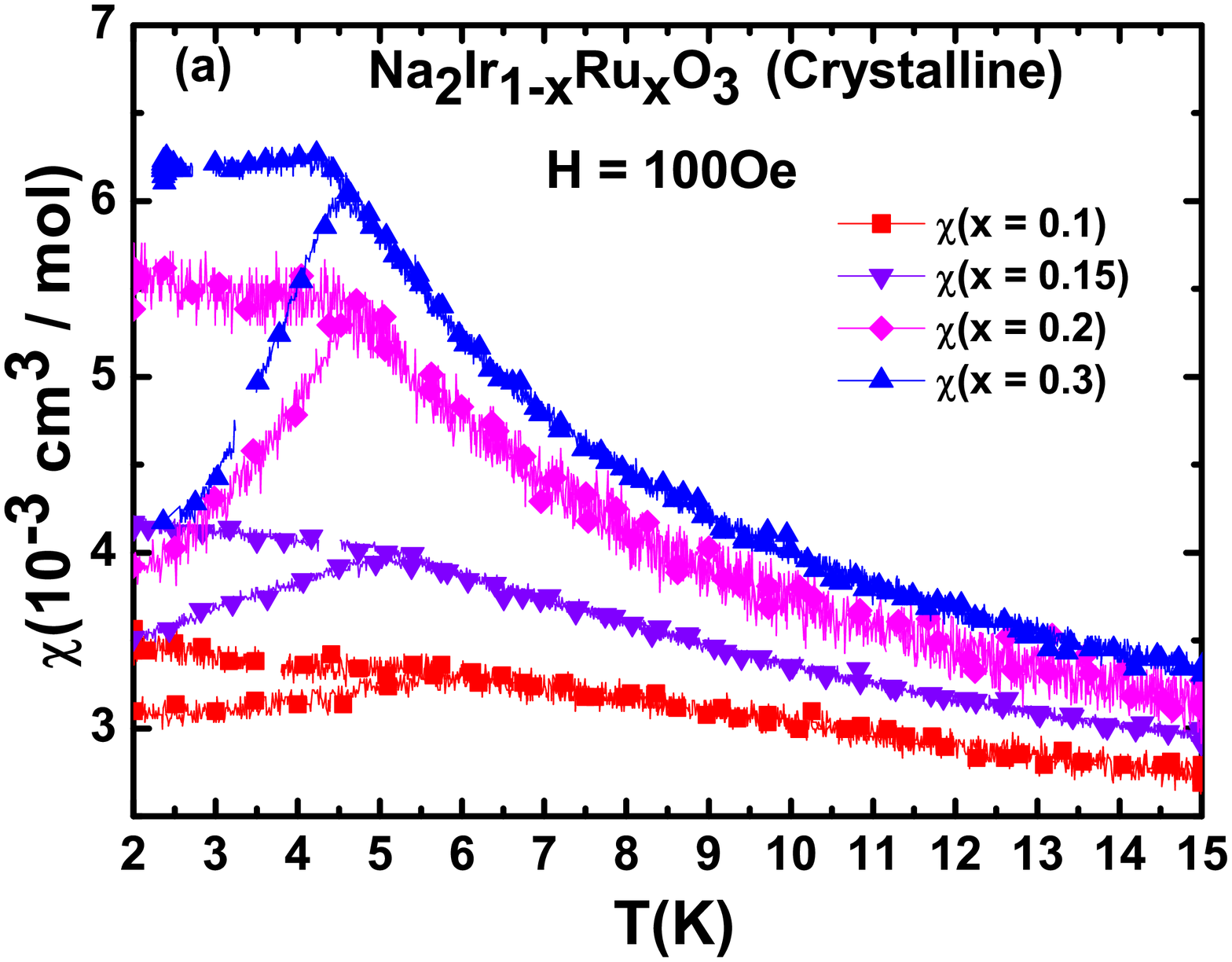}
\includegraphics[width= 3 in]{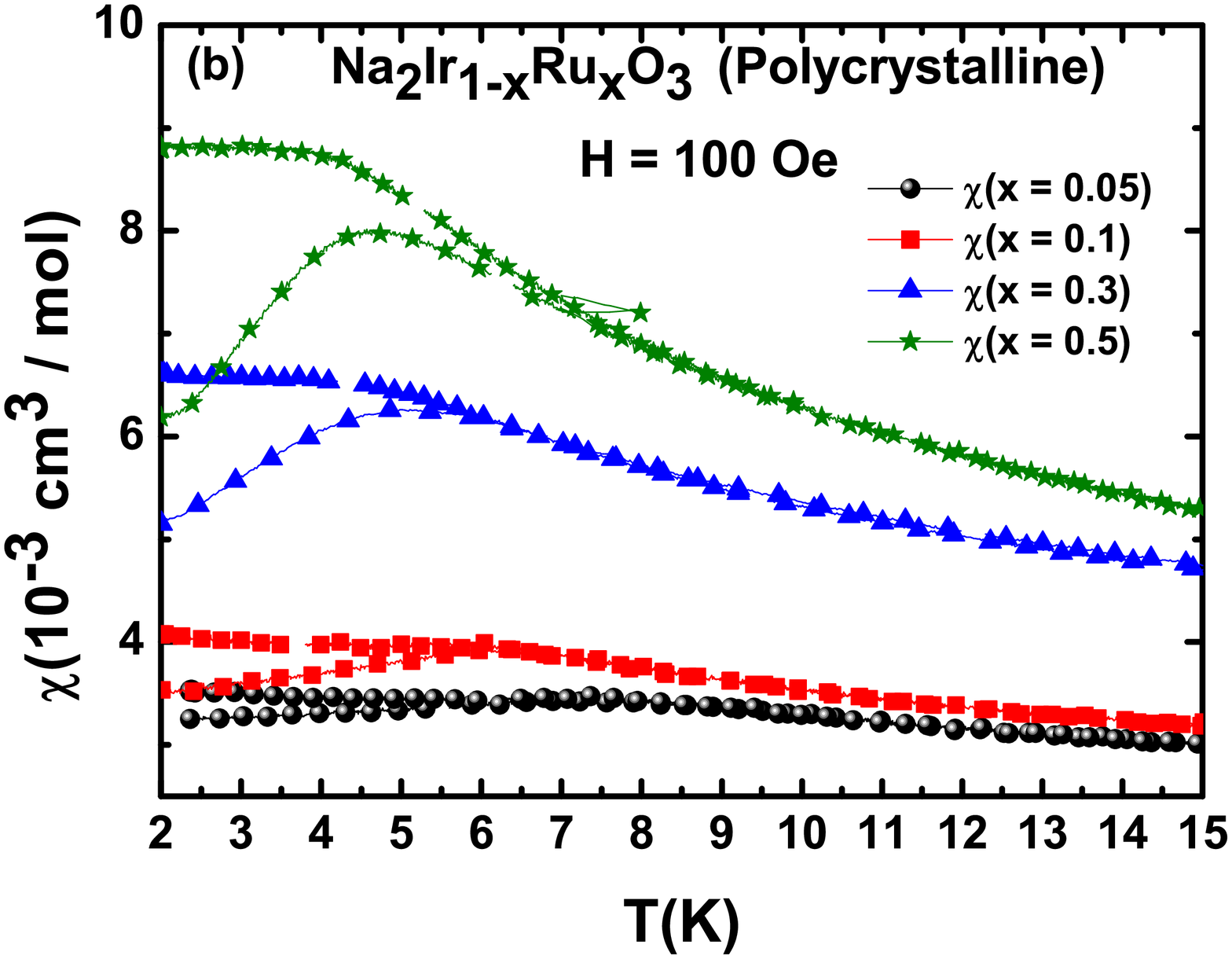}      
\caption{(Color online) Zero-field-cool (ZFC) and field-cool (FC) magnetic susceptibility $\chi$ versus $T$ curves of Na$_2$Ir$_x$Ru$_{1-x}$O$_3$ between $T = 2$~K and 15~K\@. at applied  magnetic field $ H = 100~Oe $ (a) Show ZFC - FC data of single crystalline Na$_2$Ir$_x$Ru$_{1-x}$O$_3$ ($x \approx  0.1, 0.15, 0.2, 0.3 $) at applied field $ H = 100~Oe $       (b) Show ZFC - FC data of Polycrystalline Na$_2$Ir$_x$Ru$_{1-x}$O$_3$ ($x \approx  0.05, 0.1, 0.2, 0.3 $) at applied field $ H = 100~Oe $.       
\label{Fig-2}}
\end{figure}

Magnetic irreversibility is seen at low temperatures for all Ru substituted samples.  Figures~\ref{Fig-2}~(a) and~(b) show the zero-field-cooled (ZFC) and field-cooled (FC) data between $T = 2$~K and $15$~K measured in a small magnetic field of $H = 100$~Oe for single crystalline and polycrystalline samples, respectively.  The single crystalline samples show a sharp cusp in the ZFC data around $T_g = 4.5~ \emph{--}~5.5$~K for all Ru substituted samples and there is a bifurcation between the ZFC and FC data below this temperature.  A similar behavior is observed for the polycrystalline samples also, although the cusps in the ZFC $\chi(T)$ data are not as sharp and they occur at a slightly higher temperature compared to the single crystalline samples.  The rounding of the cusp in the polycrystals maybe due to microscopic inhomogeneity in the Ru distribution.  The splitting of ZFC and FC susceptibility observed for all samples can be taken as a first indication of a frozen spin glass like state for the substituted samples.  AC susceptibility and heat capacity measurements presented below support this inference.  
 
\begin{figure}[t]
\includegraphics[width= 3 in]{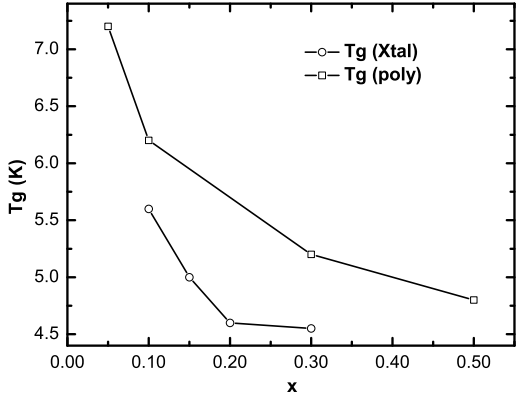} 
\caption{(Color online) The freezing temperature $T_{g}$ as a function of Ru concentration $x$ for single crystalline and polycrystalline samples of Na$_2$Ir$_x$Ru$_{1-x}$O$_3$ $(x = 0.05, 0.1, 0.15, 0.2, 0.3,0.5)$. 
\label{Tg(x)}}
\end{figure}

We track the freezing temperature $T_{\rm g}$, which we define as the peak temperature of the cusp in the DC susceptibility, with Ru concentration $x$ for single crystalline and polycrystalline samples.  These data are shown in Fig.~\ref{Tg(x)}.  It can be seen that $T_{\rm g}$ after reducing sharply for small $x$, tends to saturate to a value similar to $T_{\rm g} \approx 4.5~ \rm{to} ~4.8$~K for both kinds of samples, respectively.

\subsection{AC Magnetic Susceptibility}
\begin{figure}[t]
\includegraphics[width= 3 in]{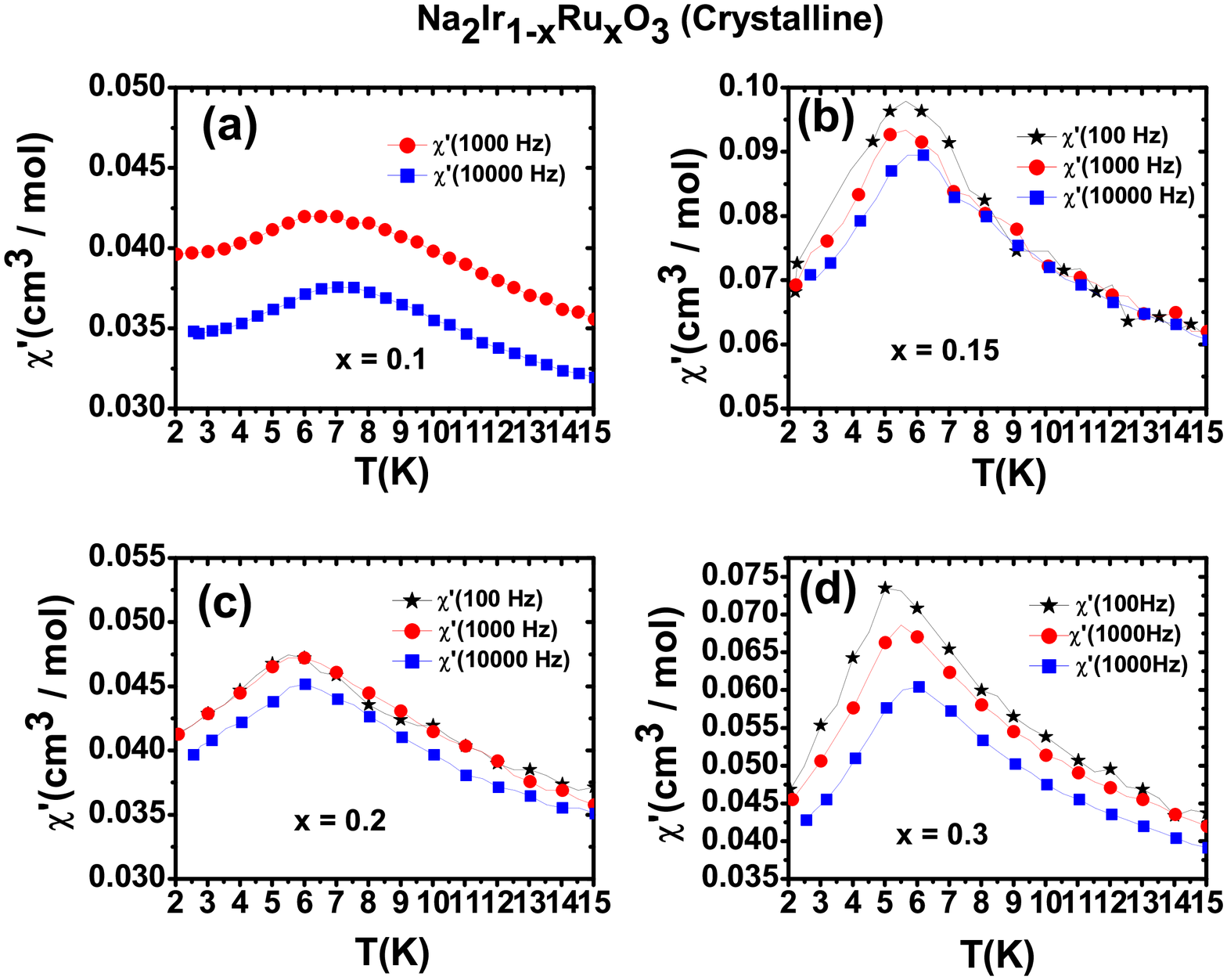} 
\caption{(Color online) The real part of ac susceptibility  $\chi'_{ac}$ as a function of temperature T = 2 K and 15 K at different frequencies  for single crystalline    Na$_2$Ir$_x$Ru$_{1-x}$O$_3$ $(x = 0.1, 0.15, 0.2, 0.3)$ at different frequency. 
\label{Fig-3}}
\end{figure}

\begin{figure}[t]
\includegraphics[width= 3 in]{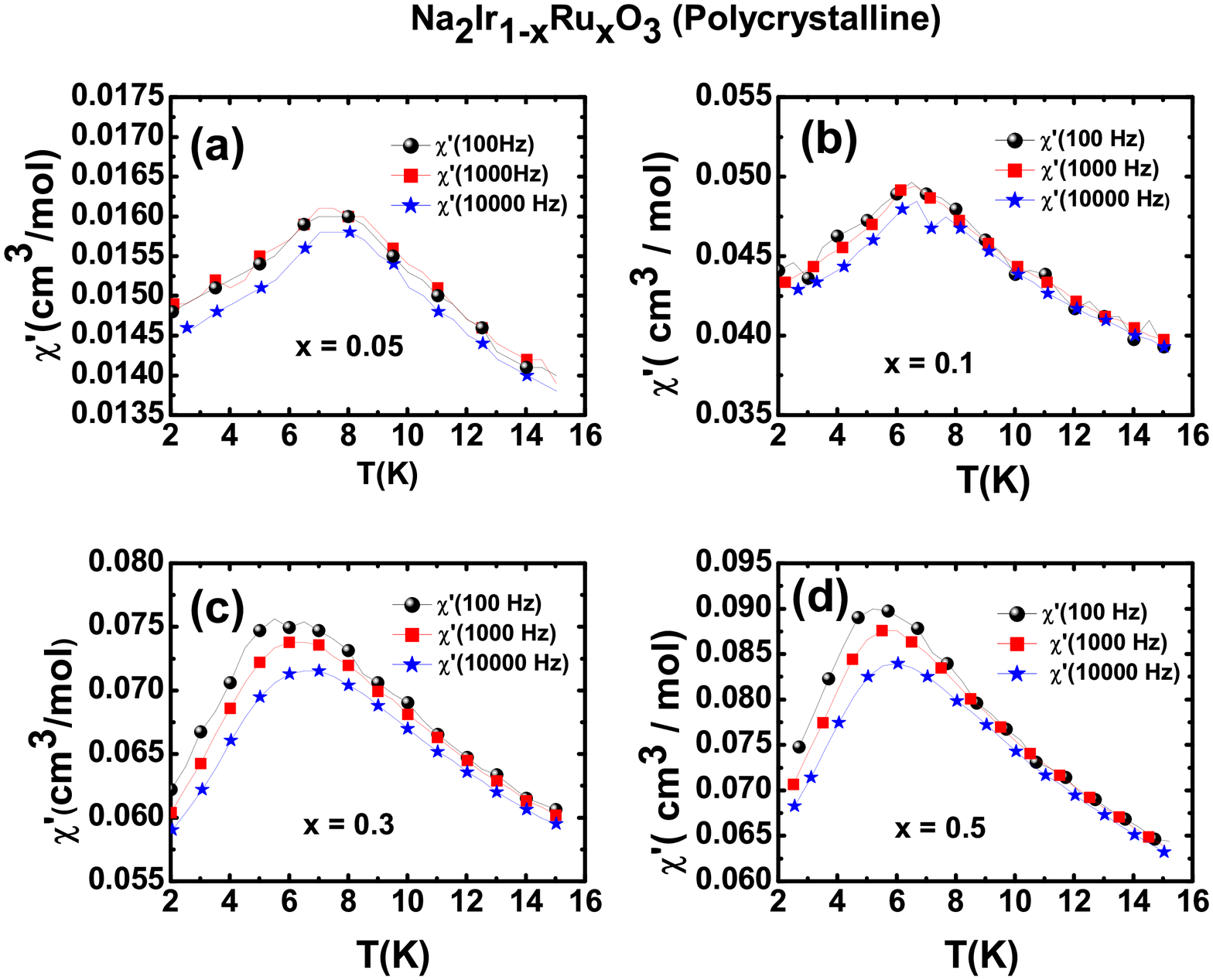}   
\caption{(Color online) The real part of ac susceptibility  $\chi'_{ac}$ as a function of temperature T = 2 K and 15 K at different frequencies  for polycrystalline Na$_2$Ir$_x$Ru$_{1-x}$O$_3$ $(x = 0.05, 0.10, 0.3, 0.5)  $ at different frequency.         
\label{Fig-4}}
\end{figure}

To check the possibility of spin glass like behavior we have measured ac susceptibility $\chi_{ac}$ at various excitation frequencies $ f $ for single crystalline and polycrystalline Na$_2$Ir$_x$Ru$_{1-x}$O$_3$ samples.  The real part of the ac susceptibility $\chi'_{ac}$ data between $T = 2$~K and $15$~K measured at various $f$ are shown in Fig.~\ref{Fig-3} and Fig.~\ref{Fig-4} for single crystalline and polycrystalline samples, respectively.  For both kinds of samples, a sharp cusp is observed at low frequency ($f = 100$~Hz) at a temperature near the cusp temperature seen in the DC susceptibility.  The position of this cusp monotonically shifts up in temperature with increasing frequency $f$ as can be seen for all samples in Fig.~\ref{Fig-3} and in Fig.~\ref{Fig-4}.  This shift of the cusp to higher temperatures with increasing frequency is a classic signature observed in canonical spin-glasses like Cu-Mn~~ \cite{Mydosh} and is strong evidence of a frozen spin-glass state below $T_g$ in our Ru substituted samples.  

A quantitative measure of the shift in the peak position with frequency is usually made using the ratio $\Delta T_{g}\over T_{g}\Delta log(f)$, where $ \Delta T_{g}$ is shift in the freezing temperature $T_{g}$ and $\Delta log(f)$ is the decade change in the frequency f.  The value of the ratio $ \Delta T_{g}\over T_{g}\Delta log(f)$ obtained for the various samples Na$_2$Ir$_x$Ru$_{1-x}$O$_3$ ($x \approx 0,0.05,0.1, 0.15, 0.2, 0.3, 0.5$) are given in Table~\ref{Table-f}. These values are typical of what has been observed for other insulating spin glass like $ Eu_{x} Sr_{1-x} S $  $ ({\Delta T_{g}\over T_{g}\Delta log(f)} \approx 0.06) $ and $ Fe_{x} Mg_{1-x} Cl_{2} $  $({\Delta T_{g}\over T_{g}\Delta log(f)} \approx 0.06) $. But these values are much larger than the value observed in canonical metallic spin glass like CuMn $({\Delta T_{g}\over T_{g}\Delta log(f)} \approx 0.005) $ ~\cite{Mulder1981}.  \

\begin{table}[t]
\caption{ Parameters obtained from the calculation for  $f$ = 100 Hz  and  10 kHz  by using the ratio                                 $ \Delta T_{g}\over T_{g}\Delta log(f)$} 

\begin{ruledtabular}
\begin{tabular}{|c|c|c|}

x & Polycrystalline ($ \Delta T_{g}\over T_{g}\Delta log(f) $) & Crystalline ($ \Delta T_{g}\over T_{g}\Delta log(f) $)\\ \hline  
0 &    &             \\
0.05 & 0.039(3)&                 \\  
0.1 & 0.014(3)&0.015(4)             \\
0.15 &    & 0.055(5)                    \\ 
0.2 &     & 0.048(1)                \\
0.3 & 0.09(1)    &  0.10(4)                 \\
0.5 & 0.16(5)&                    \\ 
\end{tabular}

\end{ruledtabular}
\label{Table-f}
\end{table}

\subsection{Heat Capacity}
\begin{figure}[t]
\includegraphics[width= 3 in]{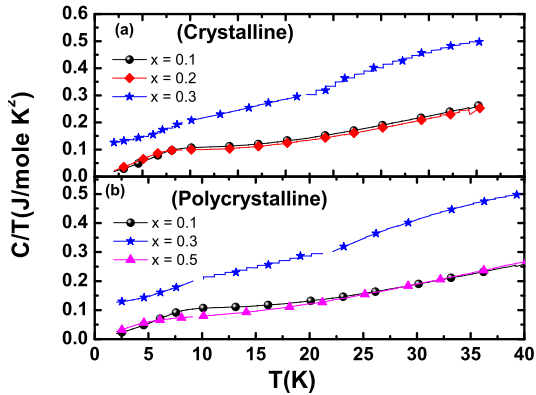}  
\caption{(Color online) (a)The heat capacity divided by temperature $C/T$ versus $T$ for crystalline Na$_2$Ir$_x$Ru$_{1-x}$O$_3$  $(x = 0.1, 0.2, 0.3)$.  (b) The heat capacity $C/T$ versus $T$ for polycrystalline Na$_2$Ir$_x$Ru$_{1-x}$O$_3$ $(x = 0.1, 0.3, 0.5)$.
\label{Fig-6}}
\end{figure}

Figure~\ref{Fig-6} shows the heat capacity divided by temperature $ C/T $ versus $ T $ data between $T = 2$~K and $T = 40$~K\@. There is no signature of any phase transition in $ C(T) $ data.  Specifically, we do not observe any sharp anomaly at the temperatures at which we observed sharp cusps in the magnetic measurements.
Instead, a broad anomaly is observed with a maximum at a higher temperature $ T \approx 10$~K\@.  These observations confirm that the anomaly at $T_{g}$ seen in $\chi (T)$ data does not arise from a bulk magnetic phase transition and provides strong evidence for a frozen spin-glass state below $T_g$ for all Ru substituted samples.

\section{SUMMARY AND DISCUSSION}
We have successfully synthesized crystalline and polycrystalline samples Na$_2$Ir$_x$Ru$_{1-x}$O$_3 (x = 0, 0.05, 0.1, 0.15, 0.2, 0.3, 0.5)$ having a honeycomb lattice of magnetic ions and have investigated their electrical and magnetic properties using electrical transport, AC and DC magnetic susceptibility, and heat capacity measurements.  All samples were found to be local moment insulators.  This is in contrast with Ru substituted Sr$_2$IrO$_4$ where an insulator to metal change is found on increasing Ru content~\cite{Qi2012} but is similar to the behavior observed for Ru substituted Li$_2$IrO$_3$~  \cite{Lee2014}.  The magnetic behavior of Ru substituted Na$_2$IrO$_3$ is however, very different from that of Ru substituted Li$_2$IrO$_3$ ~\cite{Lee2014}.  In Li$_2$IrO$_3$, the long ranged antiferromagnetic order observed at $T_N = 15$~K is suppressed to lower temperatures on increasing Ru content untill it is suppressed to below $T = 2$~K for a Ru content of $x = 0.30$  ~\cite{Lee2014}.  In contrast, for Na$_2$IrO$_3$ we find that even the smallest Ru substitution of $x = 0.05$ ($5\%$) leads to the long range antiferromagnetic order ($T_N \approx 15$~K) of the parent compound to be replaced by a frozen spin-glassy state.  Additionally, in contrast to canonical spin-glass systems like Cu-Mn where the freezing temperature roughly scales with the concentration of magnetic impurities, the freezing temperature for our Ru substituted samples decreases sharply and then settles to a constant value.  

Thus, our results suggest that the magnetic order in Na$_2$IrO$_3$ is very fragile.  Even 5\% magnetic impurity is enough to induce a frozen spin state, suggesting that disturbing the short ranged magnetic exchange pathways lead to drastic modification of the magnetic ground state.  This also highlights an important difference between the Na$_2$IrO$_3$ and Li$_2$IrO$_3$ systems.  Our results indicate that the magnetic behavior and the ground state is driven primarily by near-neighbor exchanges in Na$_2$IrO$_3$ since even a small disturbance in the Ir sublattice leads to disorder driven freezing of spins.  On the other hand previous results for Ru substitution in Li$_2$IrO$_3$ suggest that much longer ranged interactions are at play and these exchange pathways are not affected as drastically by small Ru substitutions.  Therefore, the magnetic order in Ru substituted Li$_2$IrO$_3$ survives at least upto $x = 0.2$ ($20\%$ Ru) while for Na$_2$IrO$_3$ even a $5\%$ Ru substitution leads to a frozen spin-glassy state.   

A recent study on one isolated Ru substitution ($x = 0.20$) has appeared \cite{Das2015}.  They conclude that the long-range magnetic ordering shifts from $15$~K for Na$_2$IrO$_3$ to $\approx 6$~K for Na$_2$Ir$_{0.8}$Ru$_{0.20}$.  This conclusion is based on the cusp in the DC magnetic susceptibility.  However, we have shown conclusively from magnetic irreversibility in ZFC-FC magnetization and from frequency dependence of AC susceptibility, that this cusp is a signature of a frozen spin-glass like state and not a signature of long ranged magnetic order.  

\paragraph{Acknowledgments.--} We thank the X-ray facility at IISER Mohali for powder XRD measurements.  YS acknowledges DST, India for support through Ramanujan Grant \#SR/S2/RJN-76/2010 and through DST grant \#SB/S2/CMP-001/2013.  KM acknowledges UGC-CSIR India for a fellowship.

\end{document}